\documentclass[aps,prd,twocolumn,groupedaddress,showpacs,amsmath,amssymb]{revtex4-1}

\usepackage{graphicx}
\usepackage{bm,color}

\newcommand{\be}{\begin{eqnarray}}
\newcommand{\ee}{\end{eqnarray}}
\newcommand{\nn}{\nonumber\\}

\def\tQ{\tau_{\rm Q}}
\def\teq{t_{\rm eq}}
\def\tex{t_{\rm ex}}
\def\hx{\hat{x}}
\def\hy{\hat{y}}

\begin{document}

\title{Out-of-equilibrium dynamics of multiple second-order quantum phase transitions 
in extended Bose-Hubbard model: 
Superfluid, supersolid and density wave
}
\author{Keita Shimizu$^1$}
\author{Takahiro Hirano$^1$}
\author{Jonghoon Park$^1$}
\author{Yoshihito Kuno$^2$}
\author{Ikuo Ichinose$^1$}
\affiliation{$^1$Department of Applied Physics, Nagoya Institute of Technology, Nagoya, 466-8555, Japan}
\affiliation{$^2$Department of Physics, Graduate School of Science, Kyoto University, 
Kyoto, 606-8502, Japan}

\date{\today}

\begin{abstract}
In this paper, we study the dynamics of the Bose-Hubbard model
with the nearest-neighbor repulsion by using
time-dependent Gutzwiller methods.
Near the unit filling, the phase diagram of the model contains density wave (DW),
supersolid (SS) and superfluid (SF).
The three phases are separated by two second-order phase transitions.
We study ``slow-quench'' dynamics by varying the hopping parameter
in the Hamiltonian as a function of time.
In the phase transitions from the DW to SS and from the DW to SF, we focus on 
how the SF order forms
and study scaling laws of the SF correlation length, vortex density, etc.
The results are compared with the Kibble-Zurek scaling.
On the other hand from the SF to DW, we study how the DW order evolves
with generation of the domain walls and vortices.
Measurement of first-order SF coherence reveals interesting behavior 
in the DW regime.
\end{abstract}

\maketitle


\section{Introduction}{\label{intro}}

Systems of ultra-cold atomic gases have the high versatility and controllability.
In the last decades, ultra-cold atomic gas systems play an important role
for the study on the quantum many physics as quantum simulators \cite{Nori,Cirac,coldatom1,coldatom2,coldatom3}.
In this paper, we study ultra-cold Bose gas systems as a quantum
simulator for out-of-equilibrium dynamics of many-body quantum systems. 
For a finite-temperature quench, from the view point of the cosmology,
Kibble \cite{kibble1,kibble2} studied how the system exhibits out-of-equilibrium behavior
and pointed out that the phase transitions lead to topological defects as a result of 
spontaneous symmetry breaking of continuous symmetries.
After the pioneering work by Kibble, Zurek \cite{zurek1,zurek2,zurek3} found
that a similar phenomenon is to be observed in experiments on
the condensed matter systems such as the superfluid (SF) of $^4$He.
Furthermore for the second-order phase transition, it was argued that physical
quantities satisfy some kind of scaling laws with respect to the quench time
that measures the speed of the ``slow quench".
The works by Kibble and Zurek stimulated many physicists, and there appeared
many theoretical and experimental studies to test this conjecture, which is sometimes
called Kibble-Zurek (KZ) mechanism and KZ scaling \cite{IJMPA}.
Recent experiments on ultra-cold atomic gases in a homogeneous density setup
verified the KZ scaling law for the correlation length and topological defect 
formation \cite{navon,navon2}.

Similar problem was also studied for quantum systems, i.e.,
how low-energy states evolve under a change of
the parameters in the Hamiltonian crossing a quantum phase transition (QPT),
i.e., the quantum quench \cite{Chomaz,dziarmaga,pol,Zoller,sondhi,Sonner,francuz}.
This problem has also attracted great interests. 
Experiments on behaviors of quantum systems through QPTs have been
already done using the ultra-cold atomic gases as a quantum 
simulator~\cite{Chen,Braun,Anquez,clark,cui}.

In the previous two papers \cite{SKHI,SHPKI}, we study the out-of-equilibrium
dynamics of the ultra-cold Bose atoms on a square optical lattice by using 
the Bose-Hubbard models. 
In the practical calculation, we fixed the on-site and nearest-neighbor (NN) repulsions
and varied the hopping amplitude in the Hamiltonian, and studied how the 
lowest-energy state evolves.
In Ref.~\cite{SKHI}, we investigated how the ground state evolves from the Mott
insulator to SF by means of the time-dependent Gutzwiller (GW) methods.
We first showed the behavior of the SF order parameter, and gave physical
pictures of the out-of-equilibrium behavior of the system.
We found that the physical quantities such as the correlation length of the SF, 
vortex density, etc. satisfy scaling laws and compared the obtained scaling exponents 
with the predicted values via the KZ hypothesis. 
On the other hand in Ref.~\cite{SHPKI}, we considered an extended Bose-Hubbard model,
which includes the NN repulsion.
Phase diagram has the SF, density wave (DW) and also the supersolid (SS).
For fairly weak NN repulsion, there exists a first-order phase transition directly
connecting the DW and SF phases accompanying a finite jump in physical
quantities \cite{Kimura,QMC}.
We focused on that parameter regime, and studied the quench dynamics from
the DW to SF, and vice-versa.

In this work, we consider the intermediate strength of the NN repulsion.
In this parameter regime, there exist two second-order phase transitions separating
the DW and SS, and also the SS and SF \cite{Kimura,QMC}.
Therefore, out-of-equilibrium dynamics of the multiple phase transitions
can be studied.
There are two out-of-equilibrium `impulse' regimes in the quench dynamics, 
and their locations are rather close with each other.
Then, it is interesting to see if scaling laws similar to the KZ hold or not,
how the existence of the intermediate SS changes the quench dynamics of
the DW and SF, etc~\cite{hybrid}.

This paper is organized as follows.
In Sec.~II, we introduce the extended Bose-Hubbard model on the square
lattice, and define order parameters used to distinguish various phases.
Equilibrium phase diagram obtained by the static GW methods is shown.
There are three phases, i.e., DW, SS and SF.

In Sec.~III, we show the results of the quench dynamics from the DW to SS,
and also from the DW to SF through the SS.
We study the behavior of the SF order parameter in detail and see if
scaling laws of the SF correlation length, etc, hold.
The results are compared with the KZ mechanism, and estimation of the critical
exponents is given.
We also calculate the SF correlation length in the SF regime and examine
what kind of state forms there.

In Sec.~IV, we study quench dynamics from the SF to DW through SS.
Behavior of the SF order parameter depends on the quench time $\tQ$.
For small $\tQ$ (fast quench), domain walls of finite-size DWs form and the amplitude
of the SF remains finite.
We also show that quantum vortices are bound on domain walls.
On the other hand for large $\tQ$ (slow quench), individual DW region is large.
However, the first-order correlation of the boson operator has a peculiar behavior.
Its origin is discussed.

Section V is devoted for conclusion and discussion.

\section{Extended Bose-Hubbard model and 
equilibrium phase diagram}{\label{BHM}}

We consider the two-dimensional extended Bose-Hubbard model (EBHM) 
described by the following Hamiltonian,
\be
H_{\rm EBH}&=&-J\sum_{\langle i,j \rangle}(a^\dagger_i a_j+\mbox{H.c.})
+{U \over 2}\sum_in_i(n_i-1)  \nn
&&+V\sum_{\langle i,j \rangle}n_in_j-\mu\sum_in_i,
\label{HBHM}
\ee
where $\langle i, j\rangle$ denotes a pair of NN sites of
a square lattice,
$a_i^\dagger \ (a_i)$ is the creation (annihilation) operator of boson at site $i$,
and $n_i=a^\dagger_i a_i$.
$J$ is the hopping amplitude, and $\mu$ is the chemical potential.
There are two kind of repulsions in the model, i.e., 
$U$ and $V$-terms in Eq.(\ref{HBHM}), which
describe the on-site and NN repulsions, respectively.
For $J,V<U$, the system is in the Mott insulator, whereas for $J>U,V$, the SF forms.
On the other hand for $V>J, U$, the DW order is realized.
As we see later on, there exists another phase, i.e., SS, which has both
the DW and SF orders.

In this paper, we consider the system near the unit filling 
$\rho={1 \over N_s}\sum_i\langle n_i\rangle=1$, where
$N_s$ is the number of lattice sites.
In most of the practical calculations, we set $N_s=64\times 64$ with the periodic
boundary condition.
In the previous work \cite{SHPKI}, we focused on the system near the half filling
$\rho=1/2$ and weak NN repulsion such as $V/U=0.05$, 
and studied the first-order phase transition between the DW and SF.
On the other hand in this work, we consider the near unit filling case $\rho\approx 1$ and 
relatively large $V$, and study the phase transitions including the DW,
SS and SF.

In the present work, we study quench dynamics of the system of $H_{\rm EBH}$.
To this end, we employ the time-dependent GW (tGW) 
methods~\cite{tGW1,tGW2,tGW3,tGW4,tGW5,tGW6,aoki}. 
The tGW methods approximate the Hamiltonian of the EBHM in Eq.(\ref{HBHM})
with a single-site Hamiltonian $H_i$ by introducing {\em local expectation value}
$\Psi_i=\langle a_i \rangle$,
\be 
&&H_{\rm GW}=\sum_i H_i,   \nn
&&H_i=-J\sum_{j\in i{\rm NN}}(a^\dagger_i\Psi_j+\mbox{H.c.})
+{U \over 2}n_i(n_i-1)  \nn
&&\hspace{1cm}+V\sum_{j\in i{\rm NN}}n_i\langle n_j\rangle-\mu n_i,
\label{HGW}
\ee
where $i{\rm NN}$ denotes the NN sites of site $i$ and for NN repulsion 
term the Hartree-Fock decoupling has been introduced.
To solve the quantum system $H_{\rm GW}$ in Eq.(\ref{HGW}),
we introduce the following site-factorized wave function,
\be 
|\Phi_{\rm GW}\rangle=\prod^{N_s}_i\Big(\sum^{n_c}_{n=0}
f^i_n(t)|n\rangle_i\Big),
\;\; a^\dagger_ia_i|n\rangle_i=n|n\rangle_i,
\label{GW}
\ee 
where $n_c$ is the maximum number of particle at each site, and we take
$n_c=6$ in the present work.
In terms of $\{f^i_n(t)\}$, the SF order parameter is given as,
\be
\Psi_i=\langle a_i \rangle=\sum^{n_c}_{n=1}\sqrt{n}f^{i\ast}_{n-1}f^i_n,
\label{BEC}
\ee
and $\{f^i_n(t)\}$ are determined by solving the following Schr\"{o}dinger equation
for various initial states,
\be i\hbar \partial_t |\Phi_{\rm GW}\rangle
=H_{\rm GW}(t)|\Phi_{\rm GW}\rangle.
\label{SEq}
\ee
The time dependence of $H_{\rm GW}(t)$ in Eq.(\ref{SEq}) comes from the quench
$J\to J(t)$ with fixed $U$ and $V$ as explained in the following section.
We employ the fourth-order Runge-Kutta method for study of the time evolution
in Eq.(\ref{SEq}).
In the practical calculation, we prepare 10 states as the initial state and 
study the time evolution of each state.
Physical quantities are obtained by averaging results of the time evolutions with
10 initial states.
Applicability and reliability of the GW methods are discussed rather in detail
in Ref.~\cite{SHPKI}.

In order to obtain the phase diagram, we calculate the following order parameters
to distinguish the above mentioned states, i.e., the DW, SF and SS,
\be
&& \Psi_i=\langle a_i\rangle, \;\;\; 
|\Psi|={1 \over N_s}\sum_i|\Psi_i|,  \nonumber \\
&& \Delta_{\rm DW}={1 \over N_s}\sum_i(-1)^i\langle n_i \rangle, \label{orderPs}  \\
&&\Delta_{\rm SF}={1 \over N_s}\sum_i(-1)^i|\Psi_i|, \nonumber
\ee
where $(-1)^i$ stands for $+1 \ (-1)$ for even sites (odd sites).
In Eq.(\ref{orderPs}),
$\Psi_i$ and $|\Psi|$ measure the SF order, and 
$\Delta_{\rm DW}$ for the DW,
whereas a finite $\Delta_{\rm SF}$ indicates the existence of the SS, and 
$\Delta_{\rm SF}$ is called relative order parameter \cite{iskin}.
In the study of the non-equilibrium quench dynamics, the above quantities play an 
important role and they are measured as a function of time.

\begin{figure}[t]
\centering
\begin{center}
\includegraphics[width=7cm]{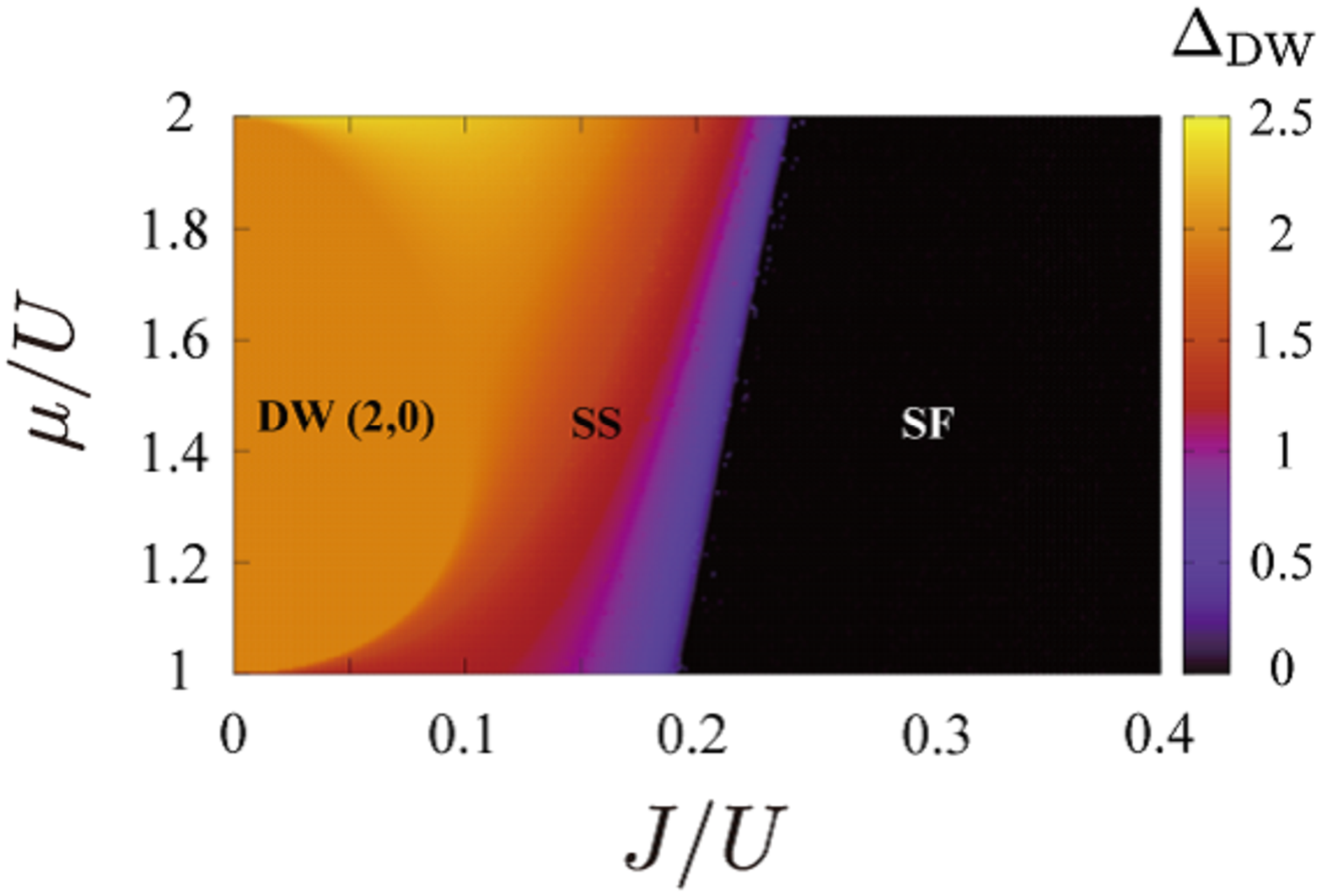}
\includegraphics[width=7cm]{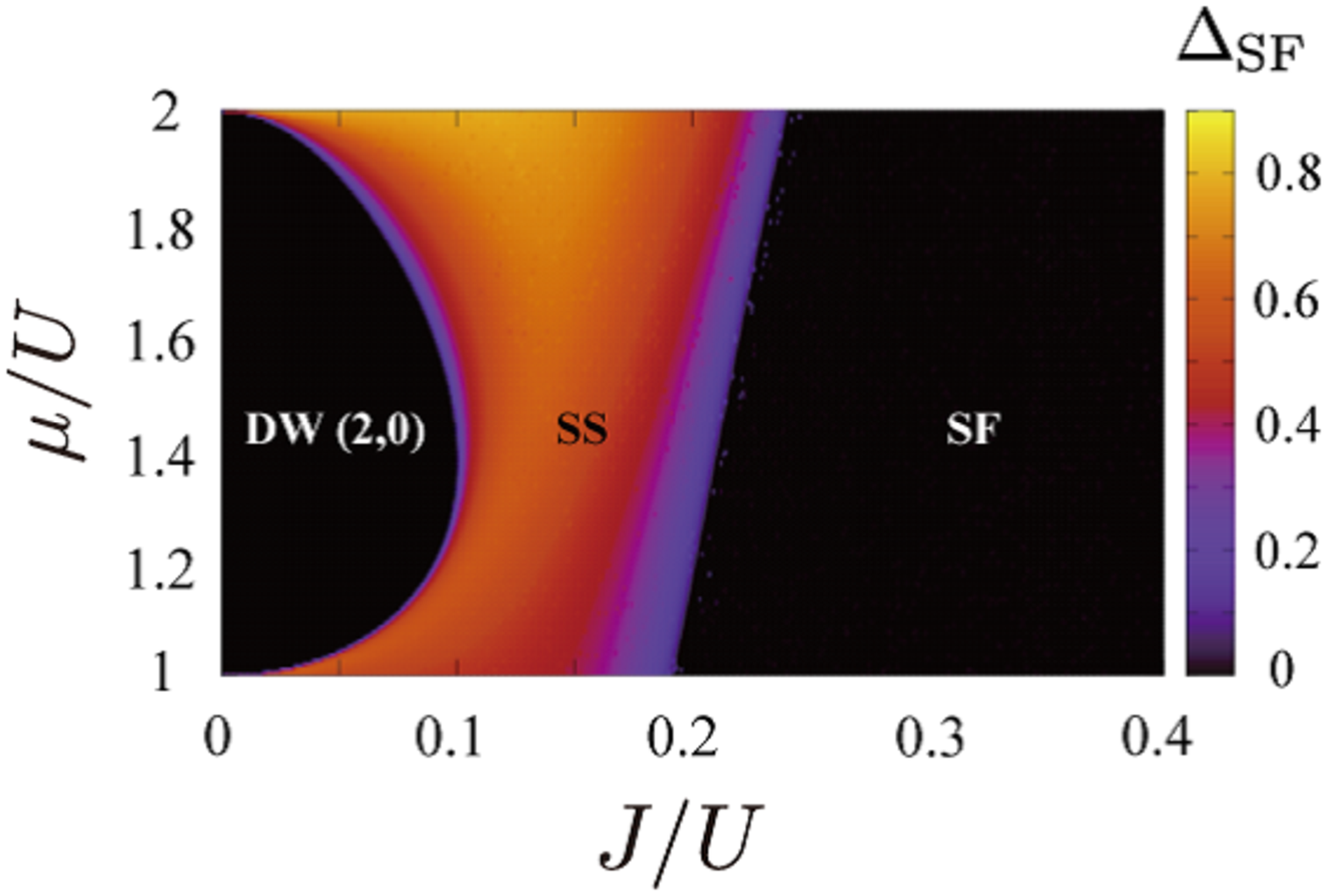}
\end{center}
\caption{Phase diagrams of the EBHM near the unit filling and $V/U=0.375$.
There are three phases, (2,0)-type DW, SS and SF.
These phases are separated by the second-order phase transitions.
In the DW, $\Delta_{\rm DW} \neq 0, \ \Delta_{\rm SF}=0, \ |\Psi|=0$.
On the other hand in the SS,  $\Delta_{\rm DW} \neq 0, \ \Delta_{\rm SF}\neq 0,
\ |\Psi|\neq 0$.
In the SF, $\Delta_{\rm DW}=0, \ \Delta_{\rm SF}=0, \ |\Psi|\neq 0$.
}
\label{phasediagram}
\end{figure}
\begin{figure}[h]
\centering
\begin{center}
\includegraphics[width=5cm]{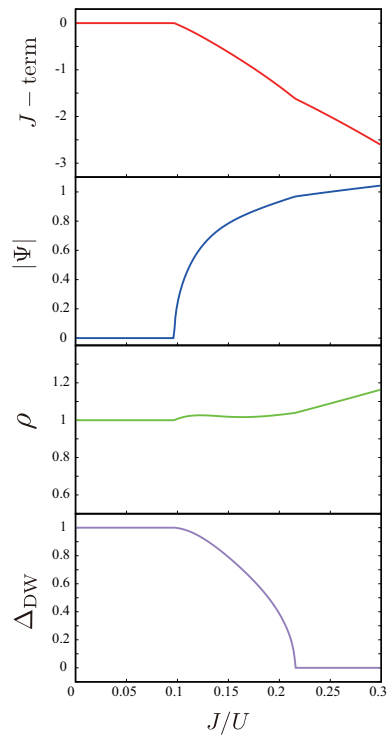}
\end{center}
\caption{Calculations of the physical quantities for the phase diagrams in
Fig.~\ref{phasediagram}.
Chemical potential $\mu/U=1.5$ and $V/U=0.375$.
$J$-term stands for the expectation value of the hopping term
$-\sum_{\langle i,j \rangle}(a^\dagger_i a_j+\mbox{H.c.})$.
$|\Psi|$ and $\Delta_{\rm DW}$ are order parameters of the SF and DW, respectively.
$\rho$ is the mean particle density.
Phase transitions take place at $J/U=J_{c1}/U\simeq 0.10$,
and $J/U=J_{c2}/U\simeq 0.22$.
}
\label{physq}
\end{figure}

Before going into the out-of-equilibrium dynamics of the system,
we show the equilibrium phase diagram of the EBHM.
To this end, we solve the time-independent Schr\"{o}dinger equation
for the Hamiltonian $H_{\rm GW}$.
We show the obtained phase diagram for $V/U=0.375$ in 
Fig.~\ref{phasediagram} and 
the physical quantities in Fig.~\ref{physq}, which are calculated by 
the static GW wave functions and used for identification of phases.
There are three phases for $V/U=0.375$, i.e., the SF, DW and SS.
The SS has both the SF and DW order, and is located between the SF and DW
in the phase diagram.
There are two phase boundaries, and both phase  transitions are of second order
as the physical quantities in Fig.~\ref{physq} indicate.
This result is in good agreement with that of the previous study using quantum
Monte-Carlo (QMC) simulations \cite{QMC} although the obtained region of the SS phase
is slightly larger than the QMC results.

In the subsequent sections, based on the phase diagram in Fig.~\ref{phasediagram},
we shall study out-of-equilibrium quench dynamics of the system that takes place 
when the system crosses the phase boundaries as a result of temporal change 
in parameters in the Hamiltonian in Eq.(\ref{HGW}).
In the practical calculation, we fix $U=1$ as the unit of energy, and also
we focus on the case with $V/U=0.375$ as in the static case.
In the previous work \cite{SHPKI}, we studied the system with $V/U=0.05$, in which 
the SS does not forms near $\rho \approx 0.5$ and 
a first-order phase boundary exists between the SF and DW.
In the present work, we are interested in how the system evolves when it
crosses the multiple second-order phase transitions, etc.

\section{Transitions from DW to SS and from DW to SF} 

\begin{figure}[h]
\centering
\begin{center}
\includegraphics[width=7cm]{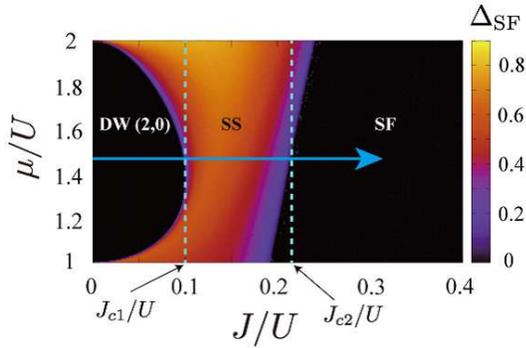}
\end{center}
\caption{Arrow indicates
quench protocol from the DW to SS and SF in the phase diagram in 
Fig.~\ref{phasediagram}.
For $\mu/U=1.5$, the critical points are located at $J_{c1}/U=0.10$
and $J_{c2}/U=0.22$.
}
\label{quench1}
\end{figure}
\begin{figure}[h]
\centering
\begin{center}
\includegraphics[width=9cm]{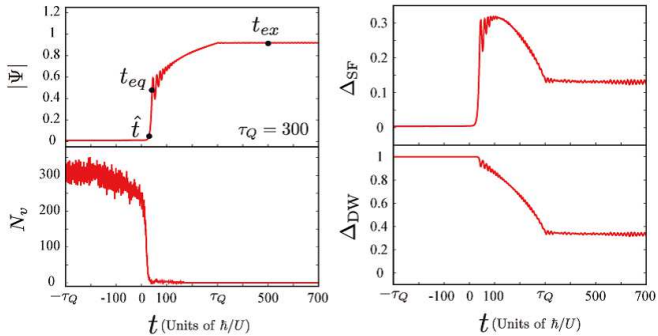}
\end{center}
\caption{Calculations of the physical quantities from the DW  to SS as a function of time.
At $t=0$, the system passes trough the equilibrium phase transition point
DW $\to$ SS.
Locations of $\hat{t}$, $\teq$ and $\tex$ are indicated.
}
\label{SFOP}
\end{figure}

In this section, we consider the dynamics of the transitions from the DW to SS 
and from the DW to SF.
As shown in Fig.~\ref{quench1}, the mean particle density $\rho\approx 1$
for $\mu/U=1.5$, and then the DW is the (2,0)-type one.
Phase transition from the DW to SS takes place at $J/U=J_{c1}/U\simeq 0.10$,
and from the SS to SF at $J/U=J_{c2}/U\simeq 0.22$, respectively.

\subsection{From DW to SS}

We study the transition from the DW to SS first.
In the practical calculation, the following quench protocol is used;
\be
{J(t)-J_{c1} \over J_{c1}}={t \over \tQ}, \; \; \; t\in [-\tQ,\tQ],
\label{proto}
\ee
where $\tQ$ is called quench time.
The protocol in Eq.(\ref{proto}) indicates that 
the system crosses the equilibrium phase transition point $J_{c1}$
at $t=0$, and the quench terminates at $t=\tQ$ with $J(\tQ)=2J_{c1}(<J_{c2})$. 

In Fig.~\ref{SFOP}, we show the order parameters 
$|\Psi|, \Delta_{\rm DW}$, and $\Delta_{\rm SF}$ as a function of 
time ($t$) for $\tQ=300$.
$|\Psi|$ exhibits a similar behavior to that in the transition from the Mott to SF
in the $V/U=0$ case studied previously~\cite{SKHI}.
As in the previous works, we define the transition time $\hat{t}$, at which
the system evolves from
the impulse to adiabatic regimes, by $|\Psi(\hat{t})|=2|\Psi(0)|$~\cite{hatt}.
On the other hand, $\teq$ is the time at which the oscillating behavior of $|\Psi|$ 
starts.
Physical picture of the oscillating regime was explained in the previous 
paper \cite{SKHI}.
The amplitude of SF, $|\Psi|$, develops quite rapidly from $\hat{t}$ to $\teq$.
On other hand, the correlation length only doubles in that period.
Genuine coarsening process of the long-range SF coherence takes place 
between $\teq$ and $\tex$, where $\tex$ is the time at which the oscillation
of $|\Psi|$ terminates. 

The other order parameters, $\Delta_{\rm DW}$, $\Delta_{\rm SF}$
and $N_v$ [defined by Eq.(\ref{vortex})]
in Fig.~\ref{SFOP}, show that the system evolves into the SS at $t=\hat{t}(=23.5)$.
This result indicates that the present definition of $\hat{t}$ is a suitable one,
that is, the adiabatic development of the SF order starts at $t=\hat{t}$.

\begin{figure}[t]
\centering
\begin{center}
\includegraphics[width=9cm]{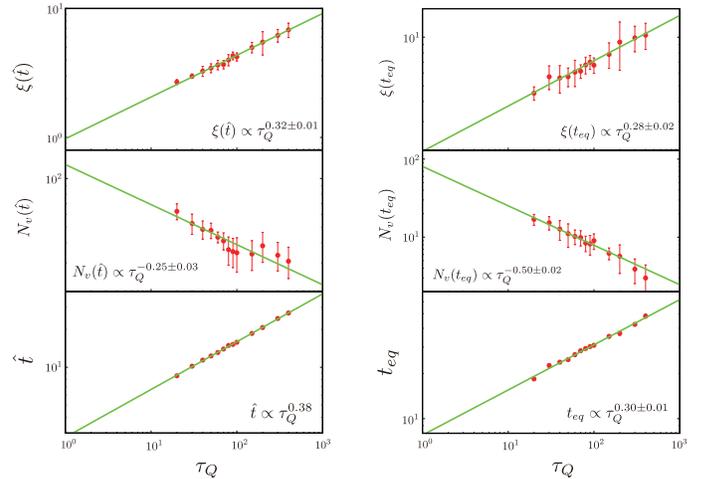}
\end{center}
\caption{Observation of scaling laws with respect to $\tQ$ for various quantities
at $t=\hat{t}$ and $t=\teq$.
The exponents are indicated with errors.
The error of the exponent in the bottom-left caption is smaller than $0.01$.
}
\label{scaling}
\end{figure}

It is quite interesting and important to see if scaling laws of physical quantities,
such as the SF correlation length and vortex density,
with respect to the quench time $\tQ$ hold or not.
Here, the SF correlation length,  $\xi$, and vortex density, $N_v$, are defined as 
\begin{eqnarray}
&&{1 \over 8N_s}\sum_{i}\langle a_i^\dagger a_{i\pm r\hat{x}(\hat{y})}
+\mbox{H.c.}\rangle 
\propto\exp (-r/\xi), \;\; (r\gg 1) \nonumber \\
&&N_{v}=\sum_i|\Omega_i|, \nonumber \\
&&\Omega_i={1 \over 4}\Big[\sin (\theta_{i+\hx}-\theta_i)
+\sin (\theta_{i+\hx+\hy}-\theta_{i+\hx})
 \nonumber \\
&&\hspace{1cm} -\sin (\theta_{i+\hx+\hy}-\theta_{i+\hy})
-\sin (\theta_{i+\hy}-\theta_{i})\Big],
\label{vortex}
\end{eqnarray}
where $\theta_i$ is the phase of $\Psi_i$
and $\hx \ (\hy)$ is the unit vector in the $x \ (y)$ direction.
As the transition from the DW to SS is of second-order, 
one may expect that the correlation length and the vortex density
satisfy a scaling law with the critical exponents of the 3D XY model, which describes 
the second-order SF phase transition.   

To see the relation between the Bose-Hubbard model and the 3D XY model, 
the path-integral quantization is useful~\cite{KI}.
By introducing the time $t$, and complex fields $\psi_i$ and
$\bar{\psi}_i$ for the operators $a_i$ and $a^\dagger_i$, respectively,
the time evolution of the system is given by the following path integral,
\be
\int [d\psi]\exp \Big[\int dt \ (-\sum_i\bar{\psi}_i\partial_t \psi_i
-iH(\bar{\psi},\psi))\Big], 
\label{Z}
\ee
where $H(\bar{\psi},\psi)$ is the Bose-Hubbard Hamiltonian with the $J$ and $U$-terms.
In the SF critical region, density fluctuations are small and the phase degrees of freedom
$\{\theta_i \}$ play an important role.
Therefore, we put $\psi_i=\sqrt{\rho_i} \ e^{i\theta_i}$, and 
expand as $\rho_i=\rho_0+\delta\rho_i$, where $\rho_0$ is the mean density
controlled by the chemical potential.
(We use the same notation $\theta_i$ as in Eq.~(\ref{vortex}), for
it essentially refers to the same thing.)
In Eq.~(\ref{Z}), 
\be
&&-\sum_i\bar{\psi}_i\partial_t \psi_i
-iH(\bar{\psi},\psi)  \nonumber \\
&&\to -i\sum_i\delta \rho_i\partial_t \theta_i+iJ\rho_0\sum_{\langle i, j \rangle}
(e^{-i\theta_i}e^{i\theta_j}+\mbox{c.c.}) \nonumber \\
&& \hspace{0.5cm}-iU\sum_i (\delta\rho_i)^2.
\label{delrho}
\ee
Integration over $\delta\rho_i$ can be readily performed as follows,
\be
&&\int d\delta\rho_i \exp\Big[-i \int dt \ (\delta \rho_i\partial_t \theta_i
+U(\delta\rho_i)^2)\Big]  \nonumber \\
&&\hspace{1cm} =e^{{i \over 4U}\int dt \ (\partial_t\theta_i)^2}.
\label{time}
\ee
Then, the resultant effective model describing the SF transition in  
the Bose-Hubbard model is given by the summation of the second $J$-term of 
Eq.~(\ref{delrho}) and the time-derivative term, 
$\sum_i(\partial_t\theta_i)^2=\sum_i(\partial_t e^{-i\theta_i}\cdot
\partial_t e^{i\theta_i})$, in Eq.~(\ref{time}).
By introducing finite slices for the time direction, the 3D XY model is realized.
The critical exponent of the spatial correlation length is given by the exponent of 
the 3D XY model, $\nu$.
Furthermore, the dynamical exponent $z=1$ as the present 3D XY model describes
(2D-space +1D-time) dynamics symmetrically, and therefore the temporal 
correlation length $\xi_t$ is proportional to the spatial correlation length $\xi$.

However, it is not obvious that the above derivation of the 3D XY model is applicable
for the present EBHM with the NN repulsion.
Furthermore, the DW and SS are not homogeneous
and also there exists the NN repulsions, and then
a simple relation between the exponents such as $d=2b$ may not hold,
where exponent $b$ for $\xi\propto \tQ^{b}$, and $d$ for $N_v\propto \tQ^{-d}$.
The above problems should be examined by the practical numerical
calculations in the present work.

We show the obtained results in Fig.~\ref{scaling} for both $t=\hat{t}$ and $t=\teq$.
It is obvious that $\xi$ and $N_v$ both satisfy a fairly good scaling law
from $\tQ=20$ to $\tQ=400$.
Exponents are estimated as $b=0.32$ and $d=0.25$ for $t=\hat{t}$,
and $b=0.28$ and $d=0.50$ for $t=\teq$, respectively.
The vortex density at $t=\teq$ is smaller compared to that at $t=\hat{t}$.
Then, the interactions between vortices are less effective at $t=\teq$, and as 
a result, the expected relation $d \approx 2b$ holds for $t=\teq$.

We also show the scaling of $\hat{t}$ and $\teq$ with respect to $\tQ$
in Fig.~\ref{scaling}.
For a second-order phase transition with the correlation-length exponent $\nu$
and dynamical exponent $z$, the KZ hypothesis predicts
$\hat{t}, \ \teq \propto \tQ^{\nu z/1+\nu z}$ and
$\xi \propto \tQ^{\nu/1+\nu z}$. 
From the above results, we can estimate the critical exponents $\nu$ and $z$ as
$\nu=0.51, \ z=1.18$ from the data at $\hat{t}$, and 
$\nu=0.40, \ z=1.07$ from the data at $\teq$, respectively.
The estimated values of $z$ are fairly close to that expected from
the 3D XY model, i.e., $z=1$.
On the other hand, the estimated values of $\nu$ do not coincide with that of the 
3D XY model, $\nu=0.672$~\cite{XYMC}.
This may imply that the DW-type inhomogeneity influences the critical behavior
of the SF order.

Finally in the above calculation, we have checked that the exponents, which
we extract, are not sensitive to the exact definition of $\hat{t}$.
In other words, in the period between $\hat{t}$ and $\teq$,  
$\xi$ and $N_v$ satisfy the scaling law quite well with the exponent close to
that of $\hat{t}$ and $\teq$.
This result implies that SF droplets develop without collapsing with each other
in that period~\cite{SKHI,SHPKI}.

\begin{figure}[t]
\centering
\begin{center}
\includegraphics[width=9cm]{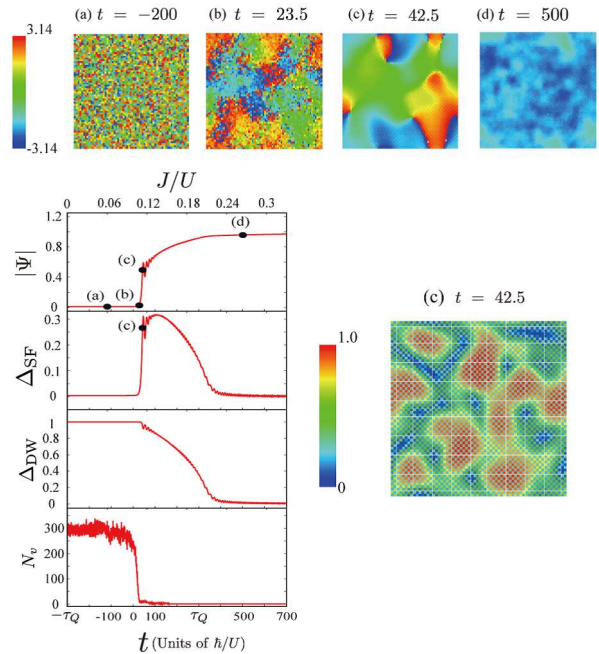}
\end{center}
\caption{Slow quench from the DW to SF through the SS.
Physical quantities as a function of time for $\tQ=300$.
$t=23.5$ corresponds to $\hat{t}$, $t=42.5$ to $\teq$ and 
$t=500$ to $\tex$.
The upper panels show phase of $\Psi_i$ for various times. 
Lower-right panel shows a profile of $|\Psi_i|$ in the SS.
The system pass through $J=J_{c1} \ (J_{c2})$ at $t=0 \ (t=1.2\tQ)$.
}
\label{DWtoSF}
\end{figure}
\begin{figure}[h]
\centering
\begin{center}
\includegraphics[width=7cm]{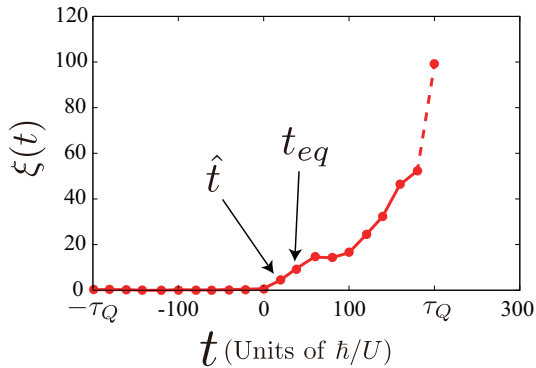}
\end{center}
\caption{SF correlation length as a function of time.
$\tQ=200$.
After $t=0$, $\xi(t)$ increases quite rapidly.
(The dotted line indicates the portion in which the correlation lengths exceed 
the system size.)
This result indicates that a SF at finite temperature forms there.
}
\label{corrL}
\end{figure}

\subsection{From DW to SF}

Let us turn to the case from the DW to SF through the SS.
Quench protocol is as follows,  
\be
{J(t)-J_{c1} \over J_{c1}}={t \over \tQ}, \; \; \; t\in [-\tQ,t_f],
\label{protoDWSF}
\ee
where we take the quench-termination time $t_f=700$ for the case of $\tQ=300$.
In Fig.~\ref{DWtoSF}, we show the behaviors of $|\Psi|$, $\Delta_{\rm DW}$,
$\Delta_{\rm SF}$ and $N_v$ as a function of time.
We also show snapshots of the phase of $\Psi_i$ in Fig.~\ref{DWtoSF} 
(the upper panels).
The DW order parameter decreases smoothly with small oscillations after the system 
passes the point $J/U\approx 0.1$, whereas the SF order parameter 
increases very rapidly after $\hat{t}$, and the coarsening process of 
the phase of $\Psi_i$ takes place smoothly from $\teq$ to $\tex$.
Order parameter $\Delta_{\rm SF}$ has nonvanishing values only in the SS.
Calculations in Fig.~\ref{DWtoSF} show that the quench dynamics from the SS 
to SF is rather smooth compared with the DW to SS.
Phase coarsening process of the SF order in the SS and SF accompanies
fluctuations of the SF amplitude as discussed in the previous work~\cite{SKHI}.
 
It is interesting to see how the correlation length evolves under the quench,
in particular, after the second critical point $J_{c2}$.
The result is shown in Fig.~\ref{corrL}.
From $\hat{t}$ to $\teq$, the correlation length doubles, whereas 
it increases rapidly after $\teq$ as a result of the coarsening process
of the phase degrees of freedom of $\Psi_i$.
The calculation in Fig.~\ref{corrL} suggests that the correlation length diverges 
for large $t$.
This result indicates that a homogeneous SF state {\em at a finite temperature} 
forms in that
regime and it has a divergent Kosterlitz-Thouless type correlation length, i.e.,
the quench of the hopping amplitude injects energy into the system, 
and an equilibrium finite-temperature SF state is realized as a result.

Here, it is suitable to comment on the definition of $\hat{t}$.
We employ its definition given in Ref.~\cite{hatt}.
Our numerical results in Fig.~\ref{DWtoSF} (in particular, the upper panels)
and Fig.~\ref{corrL} exhibit that the adiabatic development of the SF order starts 
at $t=\hat{t}$.
In fact, the phase of the SF order parameter acquires coherence at $t=\hat{t}$
as shown in Fig.~\ref{DWtoSF}.
The correlation length of the SF order also starts to increase at $t=\hat{t}$.
Therefore, it is suitable to think that a non-adiabatic chaotic state terminates
at $t=\hat{t}$.
On the other hand, the definition of $\teq$ is directly given by the behavior
of $|\Psi|$ for each $\tQ$.
In Ref.~\cite{SKHI}, we discuss that the genuine coarsening process of local
SF domains (bubbles) starts at $\teq$~\cite{SHPKI}, although
more precise study of the coarsening process is desired.

\section{Transition from SF to DW} 

\begin{figure}[h]
\centering
\begin{center}
\includegraphics[width=5cm]{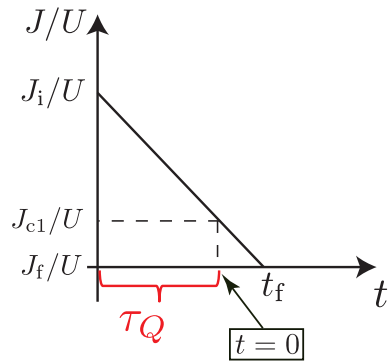}
\end{center}
\caption{Quench protocol of out-of-equilibrium dynamics in the precess
SF $\to$ SS $\to$ DW.
$(J_i/U)=0.3$ and $(J_f/U)=0$.
}
\label{protocol2}
\end{figure}
\begin{figure}[h]
\centering
\begin{center}
\includegraphics[width=8cm]{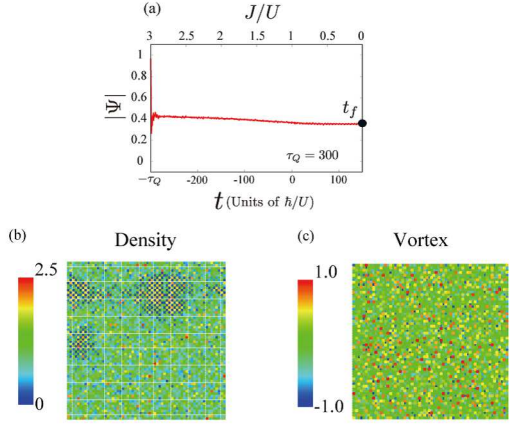}
\end{center}
\caption{If we start the time evolution with the genuine SF state with a totally
coherent phase, a DW-SF heterogeneous state forms after crossing the DW
phase transition.
We show snapshots of the heterogeneous state of the DW and SF at $t=t_f$
that forms as a result of the evolution from the genuine SF state.
Density profile exhibits clear formation of local DW regimes in the rather
homogeneous background.
Snapshot of vortices indicates that they proliferate.
}
\label{hetero}
\end{figure}

In this section, we shall study dynamical behavior of the EBHM under 
the quench from the SF to DW.
In the previous paper, we studied a related problem concerning to the
first-order phase transition from the SF to the DW~\cite{SHPKI}.
In this work, we consider the case of the multiple second-order phase transitions, i.e.,
SF $\to$ SS $\to$ DW.
As we show, the system exhibits qualitatively different behavior in the resultant
DW state depending on the value of $\tQ$.

The practical protocol is the following;
\be
{J_{c1}-J(t) \over J_{c1}}={J_i-J_{c1} \over J_{c1}}{t \over \tQ},
\; \; \; t\in [-\tQ,t_f],
\label{protoc2}
\ee
where $J_i=J(-\tQ)$ is the initial value of $J(t)$, and we choose as 
$J_i/U=0.3(>J_{c2}/U)$.
At $t=0$, $J(0)=J_{c1}$ and also we choose the final value as $J_f=J(t_f)=0$,
i.e., the quench terminates at $t=t_f={J_{c1} \over J_i-J_{c1}}\tQ$.
See Fig.~\ref{protocol2}.
As the initial state, we use a GW-type wave function, in which {\em small local
fluctuations of the phase of $\{\Psi_i\}$ are added to the equilibrium GW ground state}.
If we start the time evolution with the genuine SF state with a totally
coherent phase, a DW-SF heterogeneous state forms as we observed
in the previous work for the first-oder phase transition~\cite{SHPKI}. 
See Fig.~\ref{hetero} for the final state of the DW-SF heterogeneity.

\begin{figure}[h]
\centering
\begin{center}
\includegraphics[width=6cm]{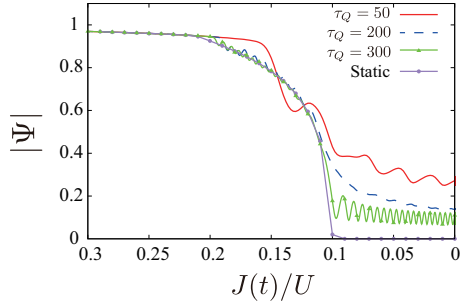}
\end{center}
\caption{SF amplitude as a function of $J/U$ for various quench times, $\tQ$'s.
Results are compared with the equilibrium values.
}
\label{SFamp2}
\end{figure}
\begin{figure}[h]
\centering
\begin{center}
\includegraphics[width=9cm]{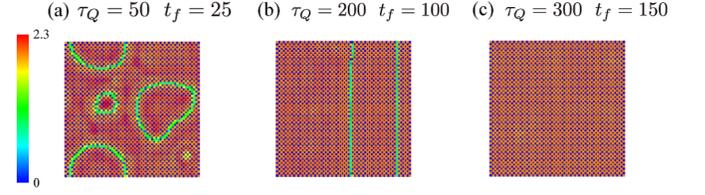}
\end{center}
\caption{Density profiles at $t=t_f$ for various quench times, $\tQ$'s.
$J(t_f)=0$.
Domain walls separating DW regions form and the total length of domain walls
decreases as $\tQ$ increases.
}
\label{density}
\end{figure}
\begin{figure}[h]
\centering
\begin{center}
\includegraphics[width=9cm]{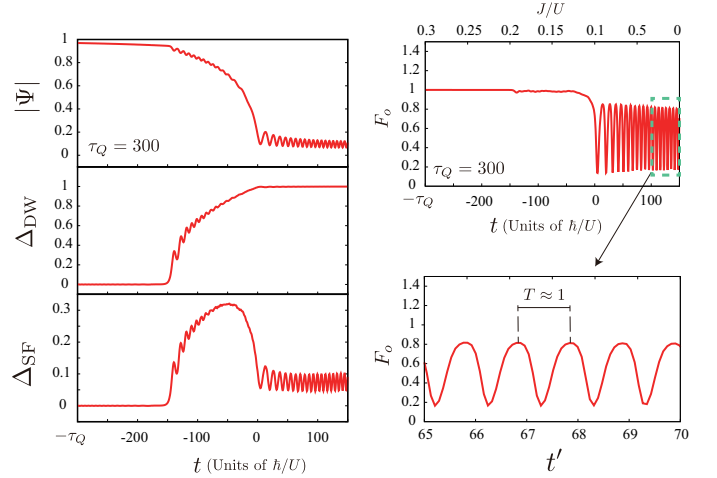}
\end{center}
\caption{Left panels:
Physical quantities $|\Psi|$, $\Delta_{\rm DW}$ and $\Delta_{\rm SF}$ 
as a function of time for $\tQ=300$.
Right panels: First-order correlation $F_o$.
After $t=0$, it exhibits the collapse-revival behavior.
$t'\equiv t\cdot {U \over 2\pi}$.
$J(0)/U=J_{c1}/U$ and $J(-150)/U\approx J_{c2}/U$.
}
\label{Pquant}
\end{figure}

We first show the SF amplitude $|\Psi|$ as a function of $J(t)/U$ in Fig.~\ref{SFamp2}
for various quench times $\tQ$'s.
As explained above, $J_{c1}/U\simeq 0.10$ and $J_{c2}/U\simeq 0.22$.
For larger $\tQ$, the results are getting closer to the static case as it is expected.
However in all cases, the SF amplitude $|\Psi|$ has a finite value for $t\to t_f$.

It is also interesting to see density profile at $t=t_f$ for the above various $\tQ$'s.
We show the obtained results in Fig.~\ref{density}.
For every $\tQ$, there are domain walls separating DW regions, and 
for larger $\tQ$, the less domain walls form.
Close look at domain walls reveals that the pattern of the DW changes as
crossing domain walls, and the expectation value of particle number at each site 
in domain walls fluctuates and takes a fractional value. 
This means that strong quantum fluctuations take place inside of domain walls.

\begin{figure*}[t]
\centering
\begin{center}
\includegraphics[width=12cm]{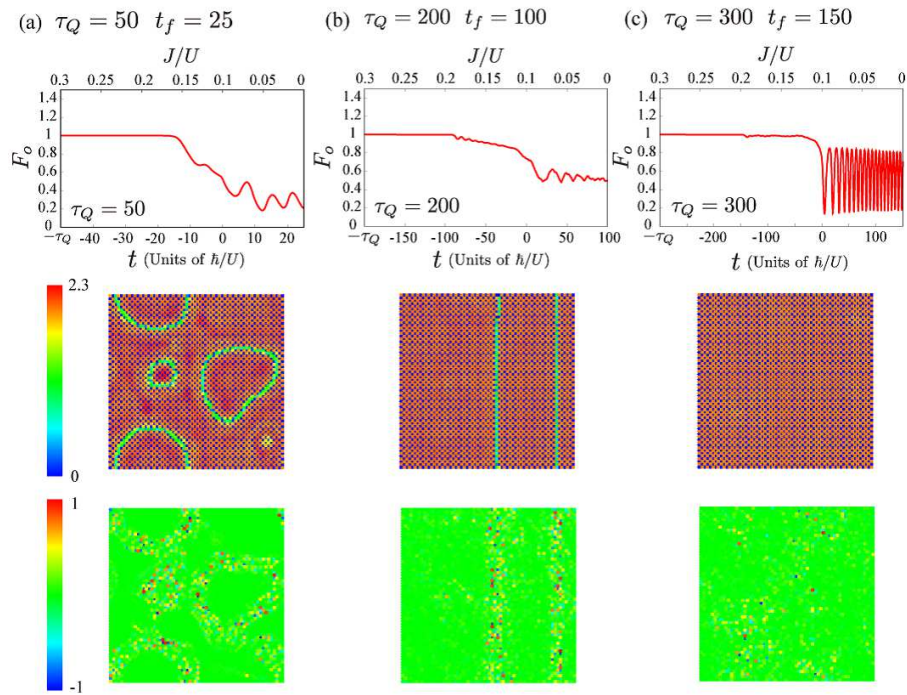}
\end{center}
\caption{Upper panel: First-order correlation function 
as a function of time for $\tQ=50, \ 200$ and $300$.
$t_f$ is the time at which the quench is terminated as shown in Fig.~\ref{protocol2}.
Middle panel: Density profiles corresponding to the above times (Fig.~\ref{density}).
Lower panel: Vortex distributions corresponding to the above times.
For $\tQ=50$ and $\tQ=200$, rather clear domain walls form although their
shapes are different in the two cases.
Vortices locate at the domain walls.
On the other hand for $\tQ=300$, locations of vortices are random.
}
\label{FoDV}
\end{figure*}

Next, physical quantities $|\Psi|$, $\Delta_{\rm DW}$ and $\Delta_{\rm SF}$ 
are shown in Fig.~\ref{Pquant} as a function of time for $\tQ=300$.
$|\Psi|$, $\Delta_{\rm DW}$ and $\Delta_{\rm SF}$ exhibit expected behaviors.
In order to investigate a SF phase coherence in detail,
we calculated the first-order correlation defined by
\be
F_o={1 \over 2N_s}\sum_{i,j}(\langle a^\dagger_i a_j\rangle +\mbox{c.c.}).
\label{FOC}
\ee
In Fig.~\ref{Pquant},
we show the calculation of $F_o$ for $\tQ=300$ as a function of time.
After $t=0$, $F_o$ exhibits fluctuating behavior and 
close look at the oscillating regime shows that the period $T\approx 1$.
In Fig.~\ref{FoDV}, we show $F_o$ as a function of time for $\tQ=50, \ 200$ and $300$.
For ever quench time $\tQ$, $F_o$ exhibits oscillating behavior after passing 
$t\approx 0$, but the pattern of oscillation strongly depends on $\tQ$.
This behavior may be related to the collapse-revival phenomenon that results
from the surviving {\em phase coherence of the SF} as studied in 
Refs.~\cite{tGW4,CR2,CR1,CR3,CR4,CR5,CR6}.
In fact for the product of the genuine coherent state, 
$|\mbox{SF}\rangle=\prod_i|\rho_i,\theta_i\rangle$
with $a_i|\rho_i,\theta_i\rangle=\sqrt{\rho_i}e^{i\theta_i}|\rho_i,\theta_i\rangle$,
$F_o$ is calculated as follows \cite{CR2},
\be
&&\langle \mbox{SF}|e^{iH_{\rm DW}t}a^\dagger_i a_je^{-iH_{\rm DW}t}
|\mbox{SF}\rangle
\nonumber \\
&&\propto \sqrt{\rho_i\rho_j} \ e^{i(\theta_j-\theta_i)}
\exp\Big\{\rho_i(e^{itU}-1)+\rho_j(e^{-itU}-1)\Big\}  \nonumber \\
&&\times \exp\Big\{\rho_k (e^{itV}-1)+\rho_\ell (e^{-itV}-1)\Big\},  
\label{FOC2}
\ee
where 
$$
H_{\rm DW}={U \over 2}\Big\{n_i(n_i-1)+n_j(n_j-1)\Big\}+V(n_in_k+n_jn_\ell).
$$
In the DW-type configurations such as $\rho_i,\rho_j \gg \rho_k,\rho_\ell$ , 
the on-site $U$-term in Eq.(\ref{FOC2}) dominates over the NN $V$-term,
and the oscillation period approximately is given by $2\pi/U$.
This explains the result in Fig.~\ref{Pquant}.

In order to verify the above expectation,
we study the cases of various $\tQ$'s, and show $F_o$ and vortex configurations 
for $J/U\approx 0$ in Fig.~\ref{FoDV}.
For $\tQ=50$ and $200$, rather clear domain walls exist, and interestingly enough,
large amount of vortices reside on these domain walls.
Therefore, the SF phase coherence is destroyed.
On the other hand for $\tQ=300$, existence of domain walls are not so clear, 
and the number of vortices is small and vortices seem locate rather randomly. 
We expect that this is the origin for the oscillating behavior of $F_o$.
In summary, we observe that for slower quench from the SF to DW, 
the SF amplitude $|\Psi|$ is getting smaller but the SF phase coherence 
is getting stronger compared to the faster quench as the vortex distribution and 
the first-order correlation $F_o$ indicate.


\section{Conclusion and discussion}

In this paper, we studied the EBHM on the square
lattice, which is expected to be realized by the ultra-cold atomic gases
and quantum simulated.
We first clarify the phase diagram of the system near the unit filling
and $V/U=0.375$.
There are three phases, the DW, SS and SF.
Then we studied the non-equilibrium quench dynamics by varying the hopping
amplitude as a function of time.

In the quench dynamics from the DW to SS, we observed the time evolution of the
SF amplitude and verified that it exhibits similar behavior in the Mott to SF 
second-order phase transition.
The correlation length of the SF order, vortex density, $\hat{t}$ and $\teq$,
all exhibit the scaling laws with respect to the quench time $\tQ$.
By using the KZ scaling hypothesis, the values of critical exponents 
$\nu$ and $z$ were estimated from our numerical simulations, and we found that
$z$ is close to the value of the 3D XY model but the estimated $\nu$ does not
agree with the value of the 3D XY model.
This discrepancy may stem from the NN repulsion and the DW order.

Next, we investigated the quench dynamics from the DW to SF through the SS.
We verified that the phase degrees of freedom of the SF order parameter experiences 
the coarsening process as in the Mott to SF transition.
The correlation length of the SF was also measured and we found that it gets large
in the SF regime.
This result implies that a SF at finite temperature forms as a result of the energy
injection by the quench.
On the other hand, the DW order smoothly decreases after passing the static
transition point to the SS and vanishes at the transition to the SF. 

Finally, we investigated the quench dynamics from the SF to DW.
The SF amplitude starts to decrease at the SF-SS transition point $J_{c2}$.
After passing the SS-DW transition point, it exhibits the oscillating behavior 
for $\tQ=300$.
Observation of the first-order correlation of the SF indicates that it is 
nothing but the collapse-revival phenomenon of the quenched SF correlation
in the DW regime.
Similar phenomenon was discussed for the SF-Mott quench dynamics in 
the previous papers \cite{SKHI,tGW4,CR2}.

We hope that the phenomena that were investigated here will be  observed
in ultra-cold atomic experiments soon.
$^{168}$Er bosonic atom is a candidate for quantum simulation of the EHBM, as
its dipole magnetic moment, $7\mu_{\rm B}$ ($\mu_{\rm B}=$ the Bohr magneton), 
is fairly large.  
In the previous paper~\cite{KSI}, we studied $^{168}$Er systems on an optical
lattice, and showed that the EBHM with $V/U \approx 0.3$ can be realized.
Furthermore, some related experiments on $^{168}$Er systems were 
performed and observation of a ground state with a DW order was reported~\cite{168Er}.

Recently, there appeared very interesting theoretical study on universality 
in the dynamics of quench phase transition~\cite{Niko}.
There, by using equations of motion or Ginzburg-Landau-type arguments, 
the KZ scaling was re-derived.
In Ref.~\cite{SHPKI}, this analysis was successfully applied to the first-order phase
transition in the EBHM in the vicinity of the half filling.
It is quite interesting to see how this approach is applied to the present
multi-second-order phase transitions.
This problem is under study, and results will be reported in a future publication.


\section*{Acknowledgments}
Y. K. acknowledges the support of a Grant-in-Aid for JSPS
Fellows (No.17J00486).



\begin{thebibliography}{99}

\bibitem{Nori}
I. M. Georgescu, S. Ashhab, and F. Nori, Rev. Mod. \\
Phys. {\bf 86}, 153 (2014).

\bibitem{Cirac}
J. I. Cirac and P. Zoller, Nat. Phys. {\bf 8}, 264 (2012).

\bibitem{coldatom1}I. Bloch, J. Dalibard, and W. Zwerger, Rev. Mod. \\
Phys. {\bf 80}, 885 (2008).

\bibitem{coldatom2}M. Lewenstein, A. Sanpera, and V. Ahufinger, 
\textit{Ultracold Atoms in Optical Lattices: 
Simulating Quantum Many-body Systems} (Oxford University Press, 
Oxford, 2012).

\bibitem{coldatom3}
K. Biedro\'{n}, M. \L{}acki, and J. Zakrzewski
Phys. Rev. B {\bf 97}, 245102 (2018).

\bibitem{kibble1}T. W. B. Kibble,
J. Phys. A: Math. Gen. {\bf 9}, 1387 (1976).

\bibitem{kibble2}T. W. B. Kibble,
Phys. Rep. {\bf 67}, 183 (1980).

\bibitem{zurek1}W. H. Zurek, 
Nature {\bf 317}, 505 (1985).

\bibitem{zurek2}W. H. Zurek, 
Acta Phys. Pol. B {\bf 24}, 1301 (1993).

\bibitem{zurek3}W. H. Zurek, 
Phys. Rep. {\bf 276}, 177 (1996).

\bibitem{IJMPA}See for example, A. del Campo and W. H. Zurek,
Int. J. Mod. Phys. A {\bf 29}, 1430018 (2014).

\bibitem{navon}N. Navon, A. L. Gaunt, R. P. Smith, and Z. Hadzibabic, \\
Science {\bf 347}, 167 (2015).

\bibitem{navon2}J. Beugnon and N. Navon, 
J. Phys. B: At. Mol. Opt. \\
Phys. {\bf 50}, 022002 (2017).

\bibitem{Chomaz}
L. Chomaz, L. Corman, T. Bienaime, R. Desbuquois, C. Weitenberg, 
S. Nascimbene, J. Beugnon, and J. Dalibard,  Nature Comm. {\bf 6}, 6172 (2015).

\bibitem{dziarmaga}J. Dziarmaga, 
Phys. Rev. Lett. {\bf 95}, 245701 (2005).

\bibitem{pol}A. Polkovnikov, 
Phys. Rev. B {\bf 72}, 161201(R) (2005).

\bibitem{Zoller}W. H. Zurek, U. Dorner, and P. Zoller,
Phys. Rev. Lett. {\bf 95}, 105701 (2005).

\bibitem{sondhi}A. Chandran, A. Erez, S. S. Gubser, and S. L. Sondhi,
Phys. Rev. B {\bf 86}, 064304 (2012).

\bibitem{Sonner}J. Sonner, A. del Campo, and W.H. Zureck,
Nature Communication {\bf 6}, 7406 (2015).

\bibitem{francuz}A. Francuz, J. Dziarmaga, B. Gardas, and W. H. Zurek,
Phys. Rev. B {\bf 93}, 075134 (2016).

\bibitem{Chen}D. Chen, M. White, C. Borries, and B. DeMarco,
Phys. Rev. Lett. {\bf 106}, 235304 (2011). 

\bibitem{Braun}S. Braun, M. Friesdrof, S. S. Hodgman, M. Schreiber, J. P. Ronzheimer,
A. Riera, M. del Rey, I. Bloch, J. Eisert, and U. Schneider,
Proc. Natl. Acad. Sci. USA {\bf 112}, 3641 (2015).

\bibitem{Anquez}M. Anquez, B. A. Robbins, H. M. Bharath, M. Boguslawski,
T. M. Hoang, and M. S. Chapman,
Phys. Rev. Lett. {\bf 116}, 155301 (2016).

\bibitem{clark}L. W. Clark, L. Feng, and C. Chin, 
Science {\bf 354}, 606 (2016).

\bibitem{cui}J-M. Cui, Y-F. Huang, Z-W. Wang, D-Y. Cao, J. Wang,
W-M. Lv, L. Luo, A. del Campo, Y-J. Han, C-F. Li, and G-C. Guo,
Sci. Rep. {\bf 6}, 33381 (2016).

\bibitem{SKHI}K. Shimizu, Y. Kuno, T. Hirano, and I. Ichinose,
Phys. Rev. A {\bf 97}, 033626 (2018).

\bibitem{SHPKI}K. Shimizu, T. Hirano, J. Park, Y. Kuno, and I. Ichinose,
New J. Phys. {\bf 20}, 083006 (2018)

\bibitem{Kimura}T. Kimura,
Phys. Rev. A {\bf 84}, 063630 (2011).

\bibitem{QMC}T. Ohgoe, T. Suzuki, and N. Kawashima,
Phys. Rev. B \\
{\bf 86}, 054520 (2012).

\bibitem{hybrid}
L-J. Zhai, H-Y. Wang, and S. Yin, Phys. Rev. B {\bf 97},  134108 (2018).

\bibitem{tGW1}
D. Jaksch, V. Venturi, J. I. Cirac, C. J. Williams, and \\
P. Zoller, 
Phys. Rev. Lett. {\bf 89}, 040402 (2002).

\bibitem{tGW2}
J. Zakrzewski, Phys. Rev. A {\bf 71}, 043601 (2005).

\bibitem{tGW3}
M. Jreissaty, J. Carrasquilla, F. A. Wolf, and M. Rigol, \\
Phys. Rev. A {\bf 84}, 043610 (2011).

\bibitem{tGW4}
M. Buchhold, U. Bissbort, S. Will, and W. Hofstetter, \\
Phys. Rev. A {\bf 84}, 023631 (2011).

\bibitem{tGW5}
S. S. Natu, K. R. A. Hazzard, and E. J. Mueller, \\
Phys. Rev. Lett. {\bf 106}, 125301 (2011).

\bibitem{tGW6}
H. Fehrmann, M. A. Baranov, B. Damski, M. Lewenstein, and L. Santos, 
Opt. Commun. {\bf 243}, 23 (2004).

\bibitem{aoki}N. Horiguchi, T. Oka, and H. Aoki,
Journal of Physics: Conference Series {\bf 150}, 032007 (2009).

\bibitem{iskin}M. Iskin, Phys. Rev. A {\bf 83}, 051606 (R) (2011).

\bibitem{hatt}P. M. Chesler, A. M. Garc\'{i}a-Garc\'{i}a, and H. Liu,
Phys. Rev. X {\bf 5}, 021015 (2015).

\bibitem{KI}Y. Kuno, K. Kataoka, and I. Ichinose,
Phys. Rev. B {\bf 87}, 014518 (2013).

\bibitem{XYMC}E. Burovski, J. Machta, N. Prokof’ev, and B. Svistunov, 
Phys. Rev. B {\bf 74}, 132502 (2006).

\bibitem{CR2}U. R. Fischer and B. Xiong, Phys. Rev. A {\bf 84}, 063635\\ 
(2011).

\bibitem{CR1}J. Dziarmaga, Adv. Phys. {\bf 59}, 1063 (2010).

\bibitem{CR3}S. Will, T. Best, U. Schneider, L. Hackermu, and I. Bloch, 
Nature {\bf 465}, 197 (2010).

\bibitem{CR4}M. Buchhold, U. Bissbort, S. Will, and W. Hofstetter, \\
Phys. Rev. A {\bf 84}, 023631 (2011).

\bibitem{CR5}C. Kollath, A. M. Lauchli, and E. Altman, 
Phys. Rev.\\ 
Lett. {\bf 98}, 180601 (2007).

\bibitem{CR6}F. Meinert, M. J. Mark, E. Kirilov, K. Lauber, P. Weinmann, 
M. Gr\"obner, and H.-C. N\"agerl, 
Phys. Rev. Lett. {\bf 112}, 193003 (2014).

\bibitem{KSI}Y. Kuno, K. Shimizu, and I. Ichinose,
Phys. Rev. A {\bf 95}, 013607 (2017).

\bibitem{168Er}S. Baier, M. J. Mark, D. Petter, K. Aikawa, L. Chomaz, Z. Cai, 
M. Baranov, P. Zoller, and F. Ferlaino, Science {\bf 352}, 201 (2016).


\bibitem{Niko}G. Nikoghosyan, R. Nigmatullin, and M. B. Plenio,
Phys. Rev. Lett. {\bf 116}, 080601 (2016).


\end{thebibliography}
\end{document}